\documentstyle[12pt]{article}

\begin{document}

\title{{\Large {\bf Geometric equations of state in Friedmann-Lema\^{i}tre
universes admitting Matter and Ricci Collineations}}}
\author{Pantelis S. Apostolopoulos and Michael Tsamparlis \and {\it \ {\small %
Department of Physics, Section of Astrophysics-Astronomy-Mechanics, }} \and 
{\it {\small University of Athens, Zografos 15783, Athens, Greece} }}
\maketitle

\begin{abstract}
As a rule in General Relativity the spacetime metric fixes the Einstein
tensor and through the Field Equations (FE) the energy-momentum tensor.
However one cannot write the FE explicitly until a class of
observers has been considered. Every class of observers defines a
decomposition of the energy-momentum tensor in terms of the dynamical
variables energy density ($\mu $), the isotropic pressure ($p$), the heat
flux $q^a$ and the traceless anisotropic pressure tensor $\pi _{ab}$. The
solution of the FE requires additional assumptions among the
dynamical variables known with the generic name equations of state. These
imply that the properties of the matter for a given class of observers
depends not only on the energy-momentum tensor but on extra a priori
assumptions which are relevant to that particular class of observers. This
makes difficult the comparison of the Physics observed by different classes
of observers for the {\it same} spacetime metric. One way to overcome this
unsatisfactory situation is to define the extra condition required among the
dynamical variables by a geometric condition, which will be based on the
metric and not to the observers. Among the possible and multiple conditions
one could use the consideration of collineations. We examine this
possibility for the Friedmann-Lema\^{i}tre-Robertson-Walker models admitting
matter and Ricci collineations and determine the equations of state for the
comoving observers. We find linear and non-linear equations of state, which
lead to solutions satisfying the energy conditions, therefore describing
physically viable cosmological models.
\end{abstract}


\section{Introduction}

In General Relativity one usually restricts the spacetime metric by means of
some symmetry conditions. The metric fixes the Einstein tensor, and through
Field Equations (FE), the energy-momentum tensor $T_{ab}$. To make Physics
one has to define physical quantities. This is done by the consideration of
a unit timelike vector field, $u^a$ say, which physically is identified with
the field of observers and geometrically it is used to 1+3 decompose $%
T_{ab}$ (and any other field) in the well known way \cite
{Wainwright-Ellis}: 
\begin{equation}
T_{ab}=\mu u_au_b+ph_{ab}+2q_{(a}u_{b)}+\pi _{ab}.  \label{sx1.1}
\end{equation}
Geometrically the quantities $\mu ,p, q_a,\pi _{ab}$ are tensors, which
depend on {\it both} $T_{ab}$ and $u^a$. Physically these quantities are the
dynamical variables (energy density, isotropic pressure, heat flux and
traceless anisotropic pressure) of the considered spacetime as observed by
the {\it specific} observers $u^a$.

The vector field $u^a$ preferably is defined from the given spacetime metric
(or energy-momentum tensor) by means of some characteristic geometric
property. For example the most standard class of observers are the comoving
observers {\em defined} by the timelike eigenvector of the Ricci (or equivalently 
the energy-momentum) tensor. For these observers $q^a =0$, that is, the
heat conduction vector vanishes\footnote{The anisotropic pressure tensor $\pi_{ab}$ 
does not necessarily vanish therefore the matter observed by the comoving observers 
is not necessarily a perfect fluid.}.
 Of course it is possible that a given
spacetime metric gives rise to more than one geometrically defined timelike
vector fields, in which case one has potentially many inherent classes of
observers.

One well known example is furnished by the spatially homogeneous
tilted perfect fluid comologies in which it is possible to consider two future directed timelike
unit vector fields \cite{King-Ellis1-1973}. The normal vector $n^a$ to the
hypersurfaces of homogeneity and the eigenvector $u^a$ of the Ricci tensor.
These vectors are not necessarily parallel and define the so called hyperbolic angle
of tilt $\beta$ by the relation $\cosh \beta=-u^an_a$. When $\beta\neq 0$ each of these vector
fields defines a class of observers. The observers $u^a$ are the comoving
ones and the vector $n^a$ is not an
eigenvector of the energy-momentum tensor and the matter
 is that of an ``imperfect
fluid with particular equations of state'' (equation (1.33a) of \cite
{King-Ellis1-1973}).

In case the metric (or any other relevant object) does not have more than
one inherent timelike vector fields, these can be introduced, if it is
considered necessary, by means of additional requirements which can be
either geometric or ``physical''.

For example the Friedmann-Lema\^{i}tre-Robertson-Walker (FLRW) model due to
its high symmetry allows only one characteristic (unit) timelike vector,
which is the timelike eigenvector of the Ricci tensor. Coley and Tupper
introduced tilted observes in flat FLRW models by the (physical) requirement
that they lead to magnetohydrodynamic viscous fluid solutions with heat
conduction \cite{Coley Tupper APJ 1983}. One can find many similar examples
in the literature.

When we write the FE in terms of the physical variables of a certain class
of observers we find a system of equations, which is not closed. This
necessitates the introduction of {\it additional} equation(s) among the
physical variables known as ``equation(s) of state'' or more general as
``constitutive equations''. For perfect fluids the standard equations
of state are relations of the form $p=p(\mu )$. It is important to note that
these additional ``contstraint'' equations are defined (a) in an a priori
manner and (b) they hold {\em only} for the specific dynamical variables,
that is, the class of observers, they refer to. Therefore the Physics of a
given spacetime of two different classes of observers cannot be compared!

One way out of this unsatisfactory situation is to consider that the
required additional equation(s) of state will be defined by means of a
geometric requirement/condition on the metric, therefore it will be common
to all classes of observers for that metric. In practice this can be
realized as follows. The condition on the metric is expressed by a set of
equations involving the metric components and possibly some other relevant
parameters. For any class of observers we express the geometric condition in
terms of the physical variables of these observers and the resulting
equations we consider to be the equation(s) of state (of {\em that} metric
for {\em these} observers).

In the case the geometric conditions are more than the required number
of equations one can define, for each appropriate subset of them, a corresponding
equation of state.

For another class of observers the {\it same} geometric condition when
written in terms of their physical variables will provide, in general,
different equation(s) of state. The important aspect is that the differences
between the equations of state will be due to the differences in the
observers only. Eventually what we have gained is:\newline
a. A way to produce equations of state consistent with the geometry of
spacetime\newline
b. A common ground on which we can compare the equations of state of
different classes of observers.\newline
One might argue that we ``let Geometry do the Physics'' and that physical
intuition is lost. But this is a common ground in General Relativity, where
we can have a ``direct'' physical picture only for the simplest cases. The
proposition we make gives more weight to the geometric consistency of the
(simplifying) assumptions involved and less to the physical intuition, and
in any case, it can be seen as a ``phenomenological'' approach, which,
however, is geometrically sound.

We have still to discuss the nature of the geometric condition on the
metric. Obviously there are many possible alternatives and, furthermore,
there is no a priori guarantee that whichever is considered will lead
to equation(s) of state, which will be of physical interest.

In this paper we propose that the condition is that the metric admits a
higher collineation (a term to be defined in section \ref{sec3}). This
proposal is logical for many reasons. Indeed one could argue that we believe
the symmetry at the level of the metric (the KVs) therefore there is no
reason why we should not believe it in its higher levels (i.e collineations).

The interplay between collineations and equations of state is not new. For
example McIntosh \cite{McIntosh1-1975} has shown that if in any spacetime,
in which we consider a class of observers such that the matter is a perfect
fluid, we require that there exists a non-trivial HVF which is normal to the
4-velocity of the observers, then the equation of state must be $p=\mu$.\footnote{%
Wainwright \cite{Wainwright1-1985} generalized this result and showed that
if in a spacetime the observers are chosen so that there is an equation of
state $p=p(\mu)$ then if we demand further that the spacetime admits a
proper HVF, the equation of state is reduced to the form $p=\alpha \mu$
where $\alpha$ is a constant.}

In the next sections we apply the above analysis to the following set:%
\newline
a. The FLRW spacetimes,\newline
b. The comoving observers for this metric i.e $u^a=\delta _0^a$,\newline
c. The geometric condition is the requirement that the metric admits a Ricci
or a matter collineation.

As it will be shown these assumptions lead to solutions of the FE, which
satisfy the energy conditions, therefore they are physically meaningful and
lead to interesting results.

The structure of the paper is as follows. In section 2 we discuss briefly
higher collineations. In section 3 we present in a concise manner the Ricci
and matter collineations and give the resulting algebraic constraints on the spatial
components of the Ricci and matter tensor. In section 4 we determine and
study the equations of state of the resulting FLRW models for the standard
comoving observers. Finally in section 5 we draw our conclusions.

\section{General comments on collineations}

\label{sec3}

A geometric symmetry or collineation is defined by a relation of the form: 
\begin{equation}
{\cal L}_{{\bf \xi }}{\bf \Phi }={\bf \Lambda }  \label{sx2.1}
\end{equation}
where $\xi ^a$ is the symmetry or collineation vector, ${\bf \Phi }$ is any
of the quantities $g_{ab},\Gamma _{bc}^a,R_{ab},R_{bcd}^a$ and geometric
objects constructed from them and ${\bf \Lambda }$ is a tensor with the same
index symmetries as ${\bf \Phi }$. By demanding specific forms for the
quantities ${\bf \Phi }$ and ${\bf \Lambda }$ one finds all the well known
collineations. For example $\Phi _{ab}=g_{ab}$ and $\Lambda _{ab}=2\psi
g_{ab}$ defines the conformal Killing vector (CKV) and specializes to a
special conformal Killing vector (SCKV) when $\psi _{;ab}=0$, to a
homothetic vector field (HVF) when $\psi =$constant and to a Killing vector
(KV) when $\psi =0.$ When $\Phi _{ab}=R_{ab}$ and $\Lambda _{ab}=2\psi
R_{ab} $ the symmetry vector $\xi ^a$ is called a Ricci conformal
collineation (RCC) and specializes to a Ricci collineation (RC) when $%
\Lambda _{ab}=0$. When $\Phi _{ab}=T_{ab}$ and $\Lambda _{ab}=2\psi T_{ab},$
where $T_{ab}$ is the energy momentum tensor, the vector $\xi ^a$ is called
a matter conformal collineation (MCC) and specializes to a matter
collineation (MC) when $\Lambda _{ab}=0$. The function $\psi $ in the case
of CKVs is called the conformal factor and in the case of conformal
collineations the {\em conformal function}.

Collineations of a different type are not necessarily independent, for
example a KV or a HVF \ is a RC\ or a MC but not the opposite. A RC or a MC
which is not a KV or a HVF (and in certain cases a SCKV) is called proper.
There are many types of collineations and most of them have been classified
in a diagram which exhibits clearly their relative properness \cite
{Katzin-Levine-Davis,Katzin-Levine}.

Although much work has been done on the computation of the higher
collineations for various types of metrics, comparatively little has been
done towards their applications in General Relativity. These applications
are usually in the direction of conservation laws. For example it has been
shown that a CKV generates a constant of motion along a null geodesic. This
result has been used to solve completely Liouville's equation for photons in
FLRW spacetimes \cite{Maartens-Maharaj,Maartens-Maharaj2}. Furthermore
timelike RCs have been related to the conservation of particle number in
FLRW spacetimes \cite{Green-Norris-Oliver-Davis}.

\section{The RCs and MCs of the FLRW models}

The geometry of the FLRW models is described by the Robertson-Walker metric
which in the standard coordinates is\footnote{%
Throughout the paper the index convention is such that latin indices $%
a,b,c,...$ take the values $0,1,2,3$ whereas Greek indices $\alpha ,\beta
,\gamma ...=1,2,3$.}: 
\begin{equation}
ds_g^2=g_{ab}dx^adx^b=S^2(\tau )\left[ -d\tau ^2+U^2(k,x^\alpha )d\sigma
_3^2\right]  \label{sx3.1}
\end{equation}
where $U(k,x^\alpha )=\left( 1+\frac k4x^\alpha x_\alpha \right) ^{-1},$ $%
k=\frac R{12},$ $d\sigma _3^2=dx^2+dy^2+dz^2$ and $R$ is the scalar
curvature of the spatial hypersurfaces. $\tau $ is the conformal time along
the world line of the comoving observers $u^a=S^{-1}\delta _\tau^a$
and it is related to the standard variable $t$ (cosmic time in FLRW
spacetime) with the relation: 
\begin{equation}
dt=S(\tau )d\tau .  \label{sx3.2}
\end{equation}
Provided the Ricci tensor $R_{ab}$ and the matter tensor $T_{ab}$ of (\ref
{sx3.1}) are non-degenerate they define the Ricci and the matter ''metric'' $%
ds_R^2,ds_T^2$ respectively by the following expressions: 
\begin{equation}
ds_R^2=R_{ab}dx^adx^b=R_0(\tau )d\tau ^2+R_1(\tau )U^2(k,x^\alpha )d\sigma
_3^2  \label{sx3.3}
\end{equation}
where: 
\begin{equation}
R_0(\tau )=\frac{3\left[ \left( S,_\tau \right) ^2-SS,_{\tau \tau }\right] }{%
S^2}  \label{sx3.4}
\end{equation}
\begin{equation}
R_1(\tau )=\frac{SS,_{\tau \tau }+\left( S,_\tau \right) ^2+2kS^2}{S^2}.
\label{sx3.5}
\end{equation}
and 
\begin{equation}
ds_T^2=T_{ab}dx^adx^b=T_0(\tau )d\tau ^2+T_1(\tau )U^2(x^\alpha )d\sigma _3^2
\label{sx3.6}
\end{equation}
where: 
\begin{equation}
T_0=3\frac{(S,_\tau )^2+kS^2}{S^2},T_1=\frac{-2SS,_{\tau \tau }+(S,_\tau
)^2-kS^2}{S^2}  \label{sx3.7}
\end{equation}
To compute the RCs and the MCs of (\ref{sx3.1}) we observe that the three
line elements $ds_g^2,ds_R^2,ds_T^2$ (not in general of the same signature!)
have the same functional form, that is, all of them can be constructed form
the conformally flat ``generic'' line element: 
\begin{equation}
ds^2=K_{ab}dx^adx^b=A^2(\tau _A)\left[ \epsilon d\tau _A^2+U^2(k,x^\alpha
)d\sigma _3^2\right]  \label{sx3.8}
\end{equation}
where the function $U(k,x^\alpha )$ has been defined above, $\epsilon =\pm 1$
for appropriate choices of the function $A(\tau _A)$ and the zero coordinate 
$\tau _A=\int \left| \frac{K_0}{K_1}\right| d\tau $. Therefore the RCs and
the MCs (for the case $R_{ab}$, $T_{ab}$ are non-degenerate\footnote{%
In case where $R_{ab},T_{ab}$ are degenerate the Lie algebra of RCs and MCs
is infinite dimensional and must be found by the solution of the FE.  
However, because they are infinite dimensional, they are not as
useful as the ones of the non-degenerate case.}) are found from the KVs of
the metric $ds^2$ provided one replaces the appropriate expressions for the
metric components $A(\tau _A).$ This has been done in \cite
{Apostolopoulos-Tsamparlis5,Apostolopoulos-Tsamparlis6} and independently in 
\cite{Camci and Barnes}. For completeness we summarize the results of these
works in Tables 1-4. The quantities $c_{\pm },\phi _k$ are defined in Table
5. \vspace{3mm}

{\footnotesize {\small TABLE 1. The proper RCs of the FLRW spacetimes for }$%
k=\pm 1${\small \ and the expression of }$R_1$ {\small for which the
corresponding collineations are admitted. }}$A${\footnotesize {\small \ is
an integration constant. Furthermore we define} $\hat{\tau}=\int \left|
R_0\right| ^{1/2}d\tau $}.

\begin{center}
\begin{tabular}{|l|l|l|}
\hline
{\bf \#} & {\bf RCs X (}$k=\pm 1)$ & ${\bf R}_1(\tau )$ \\ \hline
{\it 1} & ${\bf Y}=A\partial _{\hat{\tau}}$ & $A$ \\ \hline
{\it 4} & $
\begin{array}{c}
{\bf H}_1^k=\epsilon k\phi _k({\bf H})A\partial _{\hat{\tau}}+{\bf H}%
ta_{-\epsilon k}(\frac{\hat{\tau}}A) \\ 
{\bf Q}_\mu ^k=\epsilon k\phi _k({\bf C}_\mu )A\partial _{\hat{\tau}}+{\bf C}%
_\mu ta_{-\epsilon k}(\frac{\hat{\tau}}A)
\end{array}
$ & $\epsilon A^2c_{-\epsilon k}^2(\frac{\hat{\tau}(\tau )}A)$ \\ \hline
{\it 4} & $
\begin{array}{c}
{\bf H}_2^\epsilon =\phi _\epsilon ({\bf H})A\partial _{\hat{\tau}}-{\bf H}%
\coth (\frac{\hat{\tau}}A) \\ 
{\bf Q}_{\mu +3}^k=\phi _k({\bf C}_\mu )A\partial _{\hat{\tau}}-{\bf C}_\mu
\coth (\frac{\hat{\tau}}A)
\end{array}
$ & $\epsilon A^2\sinh ^2(\frac{\hat{\tau}(\tau )}A)$ \\ \hline
\end{tabular}
\vspace{0.1cm}
\end{center}

\clearpage

{\footnotesize {\small TABLE 2. The proper RCs of the FLRW spacetimes for }$%
k=0$. }$A${\footnotesize {\small \ is an integration constant.} }

\begin{center}
\begin{tabular}{|l|l|l|}
\hline
{\bf \#} & {\bf RCs X (}$k=0)$ & ${\bf R}_1(\tau )$ \\ \hline
{\it 4} & $
\begin{array}{c}
{\bf P}_{\hat{\tau}}=\left| A\right| ^{1/2}\partial _{\hat{\tau}} \\ 
{\bf M}_{\alpha \hat{\tau}}=x^\alpha \left| A\right| ^{1/2}\partial _{\hat{%
\tau}}-\epsilon \hat{\tau}(\tau )\left| A\right| ^{-1/2}\partial _\alpha
\end{array}
$ & $A$ \\ \hline
{\it 4} & $
\begin{array}{c}
{\bf H}=A\partial _{\hat{\tau}}+x^\alpha \partial _\alpha \\ 
{\bf K}_\alpha =2x_\alpha {\bf H}-\left( \epsilon e^{2\hat{\tau}/A}+x^\beta
x_\beta \right) \partial _\alpha
\end{array}
$ & $\epsilon A^2e^{-2\hat{\tau}(\tau )/A}$ \\ \hline
\end{tabular}
\vspace{0.1cm}
\end{center}

{\footnotesize {\small TABLE 3. The proper MCs of the FLRW spacetimes for }$%
k=\pm 1${\small \ and the expression of }$T_1$ {\small for which MCs are
admitted. }}$A${\footnotesize {\small \ is an integration constant.
Furthermore we define }$\tilde{\tau}=\int \left| T_0\right| ^{1/2}d\tau $. }

\begin{center}
\begin{tabular}{|l|l|l|}
\hline
{\bf \#} & {\bf MCs X (}$k=\pm 1)$ & ${\bf T}_1(\tau )$ \\ \hline
{\it 1} & ${\bf Y}=A\partial _{\tilde{\tau}}$ & $A$ \\ \hline
{\it 4} & $
\begin{array}{c}
{\bf H}_1^k=\epsilon k\phi _k({\bf H})A\partial _{\tilde{\tau}}+{\bf H}%
ta_{-\epsilon k}(\frac{\tilde{\tau}}A) \\ 
{\bf Q}_\mu ^k=\epsilon k\phi _k({\bf C}_\mu )A\partial _{\tilde{\tau}}+{\bf %
C}_\mu ta_{-\epsilon k}(\frac{\tilde{\tau}}A)
\end{array}
$ & $\epsilon A^2c_{-\epsilon k}^2(\frac{\tilde{\tau}(\tau )}A)$ \\ \hline
{\it 4} & $
\begin{array}{c}
{\bf H}_2^\epsilon =\phi _\epsilon ({\bf H})A\partial _{\tilde{\tau}}-{\bf H}%
\coth (\frac{\tilde{\tau}}A) \\ 
{\bf Q}_{\mu +3}^k=\phi _k({\bf C}_\mu )A\partial _{\tilde{\tau}}-{\bf C}%
_\mu \coth (\frac{\tilde{\tau}}A)
\end{array}
$ & $\epsilon A^2\sinh ^2(\frac{\tilde{\tau}(\tau )}A)$ \\ \hline
\end{tabular}
\vspace{0.1cm}
\end{center}

{\footnotesize {\small TABLE 4. The proper MCs of the FLRW spacetimes for }$%
k=0$. }$A${\footnotesize {\small \ is an integration constant.} }

\begin{center}
\begin{tabular}{|l|l|l|}
\hline
{\bf \#} & {\bf RCs X (}$k=0)$ & ${\bf T}_1(\tau )$ \\ \hline
{\it 4} & $
\begin{array}{c}
{\bf P}_{\tilde{\tau}}=\left| A\right| ^{1/2}\partial _{\tilde{\tau}} \\ 
{\bf M}_{\alpha \tilde{\tau}}=x^\alpha \left| A\right| ^{1/2}\partial _{%
\tilde{\tau}}-\epsilon \tilde{\tau}(\tau )\left| A\right| ^{-1/2}\partial
_\alpha
\end{array}
$ & $A$ \\ \hline
{\it 4} & $
\begin{array}{c}
{\bf H}=A\partial _{\tilde{\tau}}+x^\alpha \partial _\alpha \\ 
{\bf K}_\alpha =2x_\alpha {\bf H}-\left( \epsilon e^{2\tilde{\tau}%
/A}+x^\beta x_\beta \right) \partial _\alpha
\end{array}
$ & $\epsilon A^2e^{-2\tilde{\tau}(\tau )/A}$ \\ \hline
\end{tabular}
\vspace{0.1cm}
\end{center}

{\footnotesize TABLE 5. The quantities }$\phi ,c_{\pm },s_{\pm }$%
{\footnotesize \ which appear in Tables 1-4. }

\begin{center}
$
\begin{tabular}{|c|c|}
\hline
${\bf \phi }$ & ${\bf c}_{\pm }{\bf ,s}_{\pm }$ \\ \hline
$\phi _k({\bf H})=1-\frac{kU\cdot (x^\sigma x_\sigma )}2$ & $%
(c_{+},c_{-})\equiv \left[ \cosh \tilde{\tau}(\tau ),\cos \tilde{\tau}(\tau
)\right] $ \\ \hline
$\phi _k({\bf C}_\mu )=-kUx_\nu $ & $(s_{+},s_{-})\equiv \left[ \sinh \tilde{%
\tau}(\tau ),\sin \tilde{\tau}(\tau )\right] $ \\ \hline
\end{tabular}
$
\end{center}

\section{Geometric equations of state}

Consider the standard FLRW model with vanishing cosmological constant and
comoving observers $u^a=S^{-1}(\tau )\delta _\tau ^a,$ where $\tau =\int 
\frac{dt}{S(t)}$, $t$ being the standard cosmic time. For these observers
the energy-momentum tensor has a perfect fluid form i.e. $T_{ab}=\mu
u_au_b+ph_{ab}$ where $\mu ,p$ are the energy density and the isotropic
pressure measured by the observers $u^a$. This decomposition of $T_{ab}$ in
the coordinates $(\tau ,x^\alpha )$ leads to the relations: 
\begin{equation}
T_{00}=\mu S^2(\tau ),T_{11}=T_{22}=T_{33}=pS^2(\tau )U^2(k,x^\alpha ).
\label{sx4.1}
\end{equation}
Expression (\ref{sx4.1}) is a result (a) of the symmetry assumptions of the
metric and (b) the choice of observers. Using FE we compute the spatial
components of the Einstein tensor $T_{00}(S,S_{,\tau },S_{,\tau \tau
},U )$, $T_{11}(S,S_{,\tau },S_{,\tau \tau },U)$ in terms of
the scale factor $S(\tau )$ and its derivatives. From (\ref{sx3.6}), (\ref
{sx3.7}) and (\ref{sx4.1}) it follows:

\begin{equation}
\mu =3\frac{(S,_\tau )^2+kS^2}{S^4},p=\frac{-2SS,_{\tau \tau }+(S,_\tau
)^2-kS^2}{S^4}  \label{sx4.2}
\end{equation}
There remains one variable (the $S(\tau )$) free. Therefore we have to
supply one more equation in order to solve the model. This extra equation is
a barotropic equation of state, that is, a relation of the form $p=p(\mu )$.

The obvious choice is a linear equation of state $p=(\gamma -1)\mu $. There
are several solutions for this simple choice which are of cosmological
interest.\ For example $\gamma =1$ $(p=0)$ implies degeneracy of the energy
momentum-tensor (dust) and the value $\gamma =\frac 43$ ($p=\frac 13\mu )$
implies radiation dominated matter. Both states of matter are extreme and
they have been relevant at certain stages of the evolution of the Universe.
For other values of $\gamma $ one obtains intermediate states which cannot
be excluded as unphysical (see e.g. \cite{Kramer} for a thorough review).
Therefore it would be interesting to use non-linear equations of state which
will deal with more complex-and physical-forms of matter. But what will be
an ``objective'' criterion to write such equations?

We propose that this equation will be one of the constraint equations
defined by the requirement of existence of a proper RC or a MC. Of course
for every choice of observers every such equation will have a different
form, which will have to be checked that it leads to physically reasonable
results. From the geometric point of view this proposal is reasonable.
Indeed the KVs are used to fix the general form of the metric and, because a
KV is a RC and a MC, they also fix the $R_{ab},$ $T_{ab}.$ Therefore the
proper RCs and MCs are symmetries which contain the effects of covariant
differentiation ($R_{ab}$) and FE ($T_{ab}$) therefore it is reasonable to
expect that they will have immediate and stronger physical implications. One
extra advantage of this type of equations of state is that, unlikely the
standard ones, they are observer independent in the sense that they take a
specific form only after a class of observers is selected. We call these
conditions {\em geometric equations of state}.

We recall that the two important kinematic quantities in the FLRW universe are the Hubble
scalar $H$ and the deceleration parameter $q$, which are defined as
follows: 
\begin{equation}
H=\frac 13\theta =\frac{S,_\tau }{S^2}\qquad ,\qquad q=1-\frac{SS,_{\tau
\tau }}{(S,_\tau )^2}.  \label{sx4.6}
\end{equation}

In the following we demonstrate the above considerations for the case of RCs
and MCs in FLRW models and determine the equations of state for the comoving
observers. As we have mentioned in the introduction the set of equations we
shall need follows from the symmetry conditions, the FE and the conservation
equation. For the case of the FLRW cosmological models these are: 
\begin{equation}
f(\mu ,p,\mu _{,\tau },p_{,\tau })=0\hspace{2cm}\mbox{(Symmetry condition)}
\label{sx4.3}
\end{equation}
\begin{equation}
\mu _{,\tau }=-3H(\mu +p)S\hspace{2cm}\mbox{(Bianchi identity)}
\label{sx4.4}
\end{equation}
\begin{equation}
H=\frac{\sqrt{3}}3({\mu-\frac {3k}{S^2}}) ^{1/2}\hspace{2cm}\mbox{(Friedmann
equation).}  \label{sx4.5}
\end{equation}

\subsection{Geometric equations of state for MCs}

For convenience we define the new the ''time'' coordinate $\tilde{\tau}$ in
terms of the energy density of the fluid as follows: 
\begin{equation}
\tilde{\tau}(\tau )=\int \left| T_0(\tau )\right| ^{1/2}d\tau =\int \left( 3%
\frac{\left( S,_\tau \right) ^2+kS^2}{S^2}\right) ^{1/2}d\tau =\int \sqrt{%
\mu }Sd\tau .  \label{sx4.7}
\end{equation}
\underline{Case A. $k=0$}

From Table 4 we see that there are two cases to consider i.e. $T_1=A=$%
constant and $T_1=\epsilon A^2e^{-2\tilde{\tau}(\tau )/A}$. This means that
one can define two families of geometric equations of state.

\underline{Case AI: $T_1(\tau )=A\equiv \varepsilon _1a^2$ ($\varepsilon
_1=\pm 1,a\in R)$}

The constraint $T_1(\tau )=\varepsilon _1a^2$ (where $a$ is a constant)
leads to the condition: 
\begin{equation}
pS^2(\tau )=\varepsilon _1a^2  \label{sx4.8}
\end{equation}
which by means of the second of (\ref{sx4.2}) gives the equation: 
\begin{equation}
-2SS,_{\tau \tau }+(S,_\tau )^2=\varepsilon _1a^2S^2.  \label{sx4.9}
\end{equation}
The solution of (\ref{sx4.9}) provides the unknown scale factor $S(\tau )$
and describes the FLRW model completely. To solve equation (\ref{sx4.9}) we
write it in the form: 
\begin{equation}
2\left( \frac{S,_\tau }S\right) ,_\tau +\left( \frac{S,_\tau }S\right)
^2=-\varepsilon _1a^2  \label{sx4.10}
\end{equation}
which can be integrated easily. In Table 6 we give all four solutions of (%
\ref{sx4.9}) together with the physical variables of the model that is,
energy density ($\mu $), isotropic pressure ($p$), Hubble scalar ($H$) and
deceleration parameter ($q$).\newline

{\footnotesize {\small TABLE 6. The FRW models with} $k=0$ {\small which
admit the MCs }$P_{\hat{\tau}},M_{\mu \hat{\tau}}$ and $B,C$ are arbitrary
integration constants.}

\begin{center}
\begin{tabular}{|l|l|l|l|l|l|l|}
\hline
Case & $S(\tau )$ & $\mu (\tau )$ & $p(\tau )$ & $H(\tau )$ & $q(\tau )$ & 
Restrictions \\ \hline
1 & $Ce^{a\tau }$ & $-\frac{3A}{C^2e^{2a\tau }}$ & $\frac A{C^2e^{2a\tau }}$
& $\frac{3a}{Ce^{a\tau }}$ & $0$ & $\varepsilon _1=-1,A<0$ \\ \hline
2 & $B\cos ^2\frac{a\tau }2$ & $\frac{12A\left( 1-\cos a\tau \right) }{%
B^2\left( 1+\cos a\tau \right) ^3}$ & $\frac{4A}{B^2\left( 1+\cos a\tau
\right) ^2}$ & $\frac{2a\sin a\tau }{B\left( 1+\cos a\tau \right) ^2}$ & $%
\frac{1+\cos a\tau }{\sin ^2a\tau }$ & $\varepsilon _1=1,A>0$ \\ \hline
3 & $B\sinh ^2\frac{a\tau }2$ & $-\frac{3A}{B^2}\frac{\coth ^2\frac{a\tau }2%
}{\sinh ^4\frac{a\tau }2}$ & $\frac A{B^2}\sinh ^{-4}\frac{a\tau }2$ & $%
\frac aB\frac{\coth \frac{a\tau }2}{\sinh ^2\frac{a\tau }2}$ & $\frac
1{2\cosh ^2a\tau }$ & $
\begin{array}{l}
\varepsilon _1=-1,A<0 \\ 
\left( \frac{S,_\tau }S\right) ^2>a^2
\end{array}
$ \\ \hline
4 & $B\cosh ^2\frac{a\tau }2$ & $-\frac{3A}{B^2}\frac{\tanh ^2\frac{a\tau }2%
}{\cosh ^4\frac{a\tau }2}$ & $\frac A{B^2}\cosh ^{-4}\frac{a\tau }2$ & $%
\frac aB\frac{\tanh \frac{a\tau }2}{\cosh ^2\frac{a\tau }2}$ & $-\frac
1{2\sinh ^2a\tau }$ & $\left( \frac{S,_\tau }S\right) ^2<a^2$ \\ \hline
\end{tabular}
\end{center}

It is straightforward to check \ (e.g. by using an algebraic computing
program) that indeed all four solutions of FLRW\ spacetimes ($k=0$) of
Table\ 6 admit the MCs given in Table 4. A detailed study shows that all MCs
are proper, except $P_{\hat{\tau}}$ for case 1, which degenerates to a HVF.
Furthermore it can be shown that $\mu >0,\mu \pm p>0$ and $\mu +3p>0$ i.e.
all the energy conditions \cite{Hawking-Ellis} are satisfied.

Concerning the determination of the equation of state\footnote{%
Of course we could use the expressions for $S(\tau)$ and find the same
results.} we use the energy conservation equation (\ref{sx4.4}). The
symmetry condition (\ref{sx4.8}) can be written: 
\begin{equation}
2HSp=-p_{,\tau }.  \label{sx4.12}
\end{equation}
Eliminating $S(\tau )$ from (\ref{sx4.4}) and (\ref{sx4.12}) we find: 
\begin{equation}
\frac{dp}{d\mu }=\frac{p_{,\tau }}{\mu _{,\tau }}=\frac 23\frac p{p+\mu }.
\label{sx4.13}
\end{equation}
This equation has two solutions: 
\begin{equation}
p=-\frac 13\mu  \label{sx4.14}
\end{equation}
and: 
\begin{equation}
\mu -\frac{3B}a\mid p\mid ^{3/2}+3p=0.  \label{sx4.15}
\end{equation}
For $B=0$ we obtain the first solution, which is a linear equation of state
with $\gamma =\frac 23$. It corresponds to the solution of case 1 of Table 6
whose metric is: 
\begin{equation}
ds^2=-dt^2+t^{\frac 4{3\gamma }}(dx^2+dy^2+dz^2).  \label{sx4.16}
\end{equation}
This spacetime admits a HVF given by the vector $P_{\tilde{\tau}}$. The rest
three vector fields are proper MCs. We note that the spacetime (\ref{sx4.16}%
) also admits three proper RCs.

The other solution of (\ref{sx4.13}) ($B\neq 0$) leads to a non-linear
equation of state and corresponds to the cases 2,3,4 of Table\ 6.

\underline{Case AII: $T_1(\tau )=\epsilon A^2e^{-2\tilde{\tau}(\tau )/A}$}%
\newline
Using the symmetry condition $T_1(\tau )=\epsilon A^2e^{-2\tilde{\tau}(\tau
)/A}$ and equations (\ref{sx4.1}), (\ref{sx4.7}) we find: 
\begin{equation}
p=p_0S^{-3B}  \label{sx4.17}
\end{equation}
where $B=\frac 23\left( 1+\frac{\sqrt{3}}A\right) $ and $p_0=\epsilon A^2$.
Replacing this in the second equation of (\ref{sx4.2}) $(k=0)$ we get: 
\begin{equation}
2SS,_{\tau \tau }-(S,_\tau )^2=-p_0S^{4-3B}.  \label{sx4.18}
\end{equation}
Using (\ref{sx4.4}) and (\ref{sx4.17}) we can find the equation of state.
Differentiating (\ref{sx4.17}) we obtain: 
\begin{equation}
p_{,\tau }=-3BpSH  \label{sx4.19}
\end{equation}
which, when combined with (\ref{sx4.4}), gives the following equation among
the dynamic variables $\mu ,p$: 
\begin{equation}
\frac{dp}{d\mu }=\frac{Bp}{\mu +p}.  \label{sx4.20}
\end{equation}
We consider two subcases.

\underline{$B=1\Leftrightarrow A=2\sqrt{3}$}

The solution of (\ref{sx4.20}) is: 
\begin{equation}
\mu =p\mid \ln Cp\mid ,C=const,Cp>0  \label{sx4.21}
\end{equation}
which is always a non-linear equation.

\underline{$B\neq 1\Leftrightarrow A\neq 2\sqrt{3}$}

In this case the solution of (\ref{sx4.20}) is: 
\begin{equation}
p-Dp^{1/B}=(B-1)\mu ,D=const.,\,B\neq 0,1.  \label{sx4.22}
\end{equation}
Note that $D\neq 0$ is an integration constant and we have always a
non-linear barotropic equation of state.

From the above we conclude that:

{\it The only perfect fluid and flat FLRW universe with a linear equation of
state, which admits proper MCs is the FLRW model (\ref{sx4.16}) for $\gamma
=\frac 23$.}

\underline{Case B: $k=\pm 1$}

From Table 3 we observe that the possible forms of $T_1$ are $T_1=A$ and
either $T_1=\epsilon A^2c_{-\epsilon k}^2(\frac{\tilde{\tau}(\tau )}A)$ or $%
T_1=\epsilon A^2\sinh ^2(\frac{\tilde{\tau}(\tau )}A)$. For the later two
cases equations $T_{11}=T_1U^2(k,x^\mu )$ and (\ref{sx4.2}) imply:

\begin{equation}
\frac{-2SS,_{\tau \tau }+(S,_\tau )^2-kS^2}{S^2}=\epsilon A^2c_{-\epsilon
k}^2(\frac{\tilde{\tau}(\tau )}A)=pS^2  \label{sx4.23}
\end{equation}
or: 
\begin{equation}
\frac{-2SS,_{\tau \tau }+(S,_\tau )^2-kS^2}{S^2}=\epsilon A^2\sinh ^2(\frac{%
\tilde{\tau}(\tau )}A)=pS^2  \label{sx4.24}
\end{equation}
whose solution is difficult, due to high non-linear character of the above
differential equations.

For the case $T_1=A$ we obtain the equation: 
\begin{equation}
T_1=pS^2=\frac{-2SS,_{\tau \tau }+(S,_\tau )^2-kS^2}{S^2}=A.  \label{sx4.25}
\end{equation}
We observe that whenever $k\neq -A$ the resulting differential equation is
exactly the same as in the case $k=0$. Therefore the only interesting case
is when $k=-A$ which leads to the equation: 
\begin{equation}
\frac{-2SS,_{\tau \tau }+(S,_\tau )^2}{S^2}=0  \label{sx4.26}
\end{equation}
which can be written in the form: 
\begin{equation}
\frac{-2SS,_{\tau \tau }+(S,_\tau )^2}{S^2}=\frac{-2SS,_{\tau \tau
}+2(S,_\tau )^2-(S,_\tau )^2}{S^2}=-2\left( \frac{S,_\tau }S\right) _{,\tau
}-\left( \frac{S,_\tau }S\right) ^2=0.  \label{sx4.27}
\end{equation}
This equation can be solved straightforward leading to:

\begin{equation}
S(\tau )=S_0\tau ^2  \label{sx4.28}
\end{equation}
where $S_0$ is a constant of integration.

The dynamical quantities associated with the scale factor (\ref{sx4.28})
are: 
\begin{equation}
\mu =\frac{3\left( k\tau ^2+4\right) }{S_0^2t^6}  \label{sx4.29}
\end{equation}
\begin{equation}
p=-\frac k{S_0^2t^4}.  \label{sx4.30}
\end{equation}
Eliminating $t$ between $\mu ,p$ we can find the equation of state which
describe this FLRW model. Note that all the energy conditions are satisfied.

\subsection{The geometric equations of state for RCs}

The case of RCs is similar to that of MCs.  Again we distinguish cases according
to $k=0$ and $k=\pm 1$.

\underline{Case A $k=0$ $R_1(\tau )=A\equiv \varepsilon _1a^2$ ($\varepsilon
_1=\pm 1,a\in R)$}

From Table 2 it follows that we have to consider the following two cases.

\underline{Case AI $R_1(\tau )=A\equiv \varepsilon _1a^2$ ($\varepsilon
_1=\pm 1,a\in R)$}

The constraint $R_1(\tau )=\varepsilon _1a^2$ leads to the condition: 
\begin{equation}
\frac{\mu -p}2S^2(\tau )=\varepsilon _1a^2.  \label{sx4.31}
\end{equation}
Replacing in the expression of the Ricci tensor we find: 
\begin{equation}
SS,_{\tau \tau }+(S,_\tau )^2=\varepsilon _1a^2S^2.  \label{sx4.32}
\end{equation}
The solution of (\ref{sx4.32}) provides the unknown scale factor $S(\tau )$
and describes the FLRW model completely. In Table 7 we present all three
solutions of (\ref{sx4.32}).\newline

{\footnotesize {\small TABLE 7.\ The FLRW models with} $k=0$ {\small which
admit the RCs }$P_{\hat{\tau}},M_{\mu \hat{\tau}}$ and $B$ is an arbitrary integration constant.}
\begin{center}
\begin{tabular}{|l|l|l|l|}
\hline
{\bf Case} & $S(\tau )$ & $\mu (\tau )$ & {\bf Restrictions} \\ \hline
1 & $B\cos ^{1/2}a\sqrt{2}\tau $ & $\frac{3a^2\cos \sqrt{2}at\left( 1-\cos 2%
\sqrt{2}a\tau \right) }{B^2\left( 1+\cos 2\sqrt{2}a\tau \right) ^2}$ & $%
\varepsilon _1=-1$ \\ \hline
2 & $B\cosh ^{1/2}a\sqrt{2}\tau $ & $\frac{3a^2\cosh \sqrt{2}at\left( \cosh 2%
\sqrt{2}a\tau -1\right) }{B^2\left( 1+\cosh 2\sqrt{2}a\tau \right) ^2}$ & $
\begin{array}{l}
\varepsilon _1=1 \\ 
\left( \frac{S,_\tau }S\right) ^2<a^2
\end{array}
$ \\ \hline
3 & $B\sinh ^{1/2}a\sqrt{2}\tau $ & $\frac{3a^2\cosh \sqrt{2}at\sinh 2\sqrt{2%
}a\tau }{B^2\left( 1-\cosh 2\sqrt{2}a\tau \right) ^2}$ & $
\begin{array}{l}
\varepsilon _1=1 \\ 
\left( \frac{S,_\tau }S\right) ^2>a^2
\end{array}
$ \\ \hline
\end{tabular}
\end{center}

Similarly we determine the equation of state using equations (\ref{sx4.4})
and (\ref{sx4.31}). The result is:
\begin{equation}
(p-\mu )^3=C(3p+\mu )  \label{sx4.36}
\end{equation}
where $C$ is a constant of integration.

\underline{Case AII $R_1(\tau )=\epsilon A^2e^{-2\tilde{\tau}(\tau )/A}$ ($%
\epsilon =\pm 1,A\in {\bf R})$}

In this case the resulting differential equation for the determination of
the scale factor $S(\tau )$ is difficult and we have not been able to solve
it. However we can determine the equation of state. Indeed differentiating
the symmetry constraint $\left( \frac{\mu -p}2\right) S^2\equiv R_1(\tau
)=\epsilon A^2e^{-2\tilde{\tau}(\tau )/A}$ we obtain: 
\begin{equation}
\frac{\left( \mu -p\right) _{,\tau }}{\mu -p}+2\frac{S_{,\tau }}S=-\frac 2A%
\frac{d\tilde{\tau}(\tau )}{d\tau }.  \label{sx4.37}
\end{equation}
The field equations imply that $R_{ab}=(\mu +p)u_au_b+\frac{\mu -p}2g_{ab}$
therefore the $R_0-$component is $R_0=(\mu +p)S^2$. Recalling that the new
time variable $\tilde{\tau}(\tau )=\int \left| R_0\right| ^{1/2}d\tau $ we
rewrite (\ref{sx4.37}) as: 
\begin{equation}
\frac{\left( \mu -p\right) _{,\tau }}{\mu -p}+2\frac{S_{,\tau }}S=-\frac
2A(\mu +p)^{1/2}S.  \label{sx4.38}
\end{equation}
Using (\ref{sx4.4}) and (\ref{sx4.5}) in (\ref{sx4.38}) we get: 
\begin{equation}
\frac{dp}{d\mu }=\frac{\mu +5p}{3(\mu +p)}+\frac{2\sqrt{3}}3\frac{p-\mu }{%
\left[ \mu (\mu +p)\right] ^{1/2}}.  \label{sx4.39}
\end{equation}
The solution of equation (\ref{sx4.39}) in implicit form is: 
\begin{equation}
\begin{tabular}{l}
$(p-\mu )^3\left( \frac{2\sqrt{3}\sqrt{\mu (\mu +p)}+\mu +3p}{11\mu ^2+6\mu
p-9p^2}\right) ^4\left( \frac{\sqrt{3}\sqrt{\mu ^{-1}(\mu +p)}+\sqrt{3}-1}{%
\sqrt{3}\sqrt{\mu ^{-1}(\mu +p)}-\sqrt{3}-1}\right) ^{7\sqrt{3}/3}\times $
\\ 
$\hspace{1cm}\times \left[ \frac{\mu \left( \sqrt{\mu ^{-1}(\mu +p)}-\sqrt{2}%
\right) ^2}{p-\mu }\right] ^{3\sqrt{6}/2}=\mu _0$%
\end{tabular}
\label{sx4.40}
\end{equation}
where $\mu _0$ is an integration constant.

\underline{Case B $k=\pm 1$}

Again there are three cases to consider of which we have been able to solve only 
the case  $R_1=A$. In this case equation (\ref{sx3.5}) gives: 
\begin{equation}
R_1=\frac{SS,_{\tau \tau }+(S,_\tau )^2+2kS^2}{S^2}=A  \label{sx4.41}
\end{equation}
where again $A$ is an arbitrary constant.

As before the only interesting case is when $k=A$ which leads to the
equation: 
\begin{equation}
\frac{SS,_{\tau \tau }+(S,_\tau )^2}{S^2}=0.  \label{sx4.42}
\end{equation}
The solution of this equation is:

\begin{equation}
S(\tau )=S_0\tau ^{1/2}  \label{sx4.43}
\end{equation}
where $S_0$ is a constant of integration.

The dynamical quantities associated with the scale factor (\ref{sx4.43})
are: 
\begin{equation}
\mu =\frac{3\left( 4k\tau ^2+1\right) }{4S_0^2t^3}  \label{sx4.44}
\end{equation}
\begin{equation}
p=\frac{3-4k\tau ^2}{4S_0^2t^3}.  \label{sx4.45}
\end{equation}

If desired, one can compute the equation of state from (\ref{sx4.44}) and (\ref{sx4.45}).  

\section{Discussion}

In general an equation of state requires the following:

1. A metric, which leads to a given Einstein tensor.

2. A class of observers, which define the physical variables for the given
Einstein tensor

3. A number of a priori assumptions among the physical variables defined in
step 2.

We have proposed that the last step shall be replaced by a geometric
equation at the level of the metric (or any other appropriate geometric object),
 so that the equation(s) of state for a
given metric and a given class of observers will be compatible with the
geometric structure of spacetime and will follow in a systematic way from a
common assumption. This makes possible the comparison of the equations of
state, and consequently the Physics, of different classes of observers in a given  
spacetime. We
have applied this proposal to the highly symmetric FLRW spacetime for the comoving  
observers the extra geometric assumptions being that the metric admits a RC or a MC. We  
have obtained linear and non-linear equations of state,  which have a sound physical  
meaning, in the sense that the resulting models satisfy the basic requirements of a   
viable cosmological model.

\end{document}